\documentclass[sigconf,nonacm]{acmart}

\setcitestyle{numbers,sort&compress}

\AtBeginDocument{%
  \providecommand\BibTeX{{%
    \normalfont B\kern-0.5em{\scshape i\kern-0.25em b}\kern-0.8em\TeX}}}

\setcopyright{acmlicensed}
\copyrightyear{2026}
\acmYear{2026}
\acmDOI{XXXXXXX.XXXXXXX}

\acmConference[RecSys 2026]{20th ACM Conference on Recommender Systems}{Sep 28--Oct 2, 2026}{Minneapolis, Minnesota, USA}
\acmISBN{978-1-4503-XXXX-X/18/06}

\graphicspath{{./images/}} 

\usepackage{microtype}
\usepackage{graphicx}
\usepackage{booktabs} 
\usepackage{amsfonts}
\usepackage{xcolor}
\newcommand{\algcomment}[1]{\textcolor{green!40!black}{\textit{// #1}}}

\usepackage{caption}
\usepackage{subcaption}
\usepackage{tikz}
\usepackage{multirow}
\usepackage{array}
\usepackage{enumitem}

\usepackage{algorithm}
\usepackage{algpseudocode}
\usepackage{tabularx}
\usepackage{array}
\usepackage{multirow}

\usepackage[subtle]{savetrees}

\setcopyright{none}

\acmDOI{}
\acmISBN{}

\newcolumntype{Y}{>{\raggedright\arraybackslash}X}

\begin{document}

\title[\textsc{HARNESS}-LM: A Training Recipe for Sponsored Search Retrieval]{\textsc{HARNESS}-LM: A Three-Phase Training Recipe for Harnessing SLMs in Sponsored Search Retrieval}








\author{Vipul Gupta, Shikhar Mohan, Lakshya Kumar, \\
Pranjal Chitale, Nikit Begwani, Amit Singh, Manik Varma}
\affiliation{%
  \institution{Microsoft AI}
  \country{India}
}

\renewcommand{\shortauthors}{Gupta et al.}

\begin{abstract}
In the competitive landscape of sponsored search, balancing retrieval quality with production latency is a critical challenge. While large retrieval models based on Small Language Models (SLMs) such as Qwen3-Embedding-4B/8B set strong upper bounds on public benchmarks, their deployment in high-throughput, latency-sensitive environments remains impractical. In this paper, we present \textsc{Harness}-LM (HLM), a three-phase training framework for transferring the capabilities of large-scale retrievers into compact, cost-efficient models. The approach comprises: (1) training a high-performance reference (“teacher”) retriever by fine-tuning a billion-parameter-scale SLM; (2) aligning query representations via an $L_2$ objective to distill knowledge into a sub-600M parameter student encoder; and (3) applying a final contrastive refinement stage to optimize the student for retrieval performance. We also present a comprehensive empirical study of key design choices, including alignment objectives, embedding dimensionality, model scale, architecture, and optimization strategies, to identify configurations that are most effective in production settings. On a real-world Bing Ads evaluation benchmark, HLM recovers over $98\%$ of the reference retriever’s precision across multiple settings, while delivering up to $27\times$ lower online query-encoder latency and $20\times$ higher throughput on NVIDIA A100 GPUs. Online A/B testing on Bing Ads further shows a $\mathbf{+1\%}$ \textbf{Revenue}, $\mathbf{+0.6\%}$ \textbf{Impression} and $\mathbf{+0.4\%}$  \textbf{Click} uplift over current ensemble of retrievers running in production with the deployed 190M parameter model, clearly highlighting the practical efficacy of the HLM recipe in a real world sponsored search setting.
\end{abstract}



\keywords{
Dense Retrieval, LLMs, SLMs, Knowledge Distillation, Pruning, Sponsored Search, Unsupervised learning, Contrastive training
}


\maketitle

\section{Introduction}
Sponsored search is a key revenue model for search engines, where paid advertisements (Ads) are displayed alongside organic results in response to user queries. Advertisers bid on Ads, and these are ranked using multiple signals including bid value and relevance to the user’s intent. This lets search engines monetize high-intent traffic while delivering useful, contextually relevant Ads to users. Sponsored search systems rely on a high-throughput first-stage retriever to select a small set of advertisements from a very large corpus. This stage directly shapes downstream ranking quality and business metrics while operating under strict latency, throughput, and cost constraints. Dense retrieval with dual encoders~\cite{karpukhin2020dpr} is a natural fit for this setting: document embeddings can be precomputed offline and served through Approximate Nearest Neighbor (ANN)~\cite{jayaram2019diskann} search, while only the query encoder runs online. However, this deployment asymmetry also creates the central bottleneck for modern neural retrieval: the online query encoder is invoked in real-time for each query, so model size, inference latency, and serving costs tightly limit what can be deployed.

Recent Small Language Model (SLM)-based embedding models have substantially advanced text retrieval. Works such as RepLLaMA \cite{ma2024finetuning}, LLM2Vec \cite{behnamghader2024llm2vec}, NV-Embed \cite{lee2025nvembed}, GRIT-LM \cite{muennighoff2025grit}, Qwen3-Embedding \cite{zhang2025qwen3}, KaLM-Embedding-V2 \cite{zhao2025kalm}, Llama-Embed-Nemotron-8B \cite{llama_nemotron_8B}, and EmbeddingGemma \cite{vera2025embeddinggemma} demonstrate that decoder-only foundation models can be adapted into strong retrievers, achieving state-of-the-art results on multilingual benchmarks like MMTEB \cite{mmteb}. These advances suggest that sponsored search can benefit from SLMs' stronger semantic representations.

However, deploying such retrievers in real-time sponsored search is impractical due to their scale (billions of parameters), which imposes substantial GPU requirements and makes it challenging to meet stringent latency constraints ({<}15\,ms) at web scale.\footnote{SLMs are ``small'' relative to frontier Large Language Models (LLMs), but still much larger than conventional dense retrieval encoders, making online serving costly.} At the same time, these models provide a clear quality upper bound\footnote{We use ``upper-bound'' to denote the strongest retriever trainable within our compute budget; it is not a theoretical bound.} and can act as effective teachers, exploiting greater capacity and richer offline signals, but cannot serve the latency-critical online query path directly. This creates a gap between \emph{what performs best offline} and \emph{what can be served online}.

A natural solution is to fine-tune a smaller query encoder (deployed online) directly along with a larger document encoder (indexed offline), but this does not fully address the gap.
In this asymmetric setup, the online query encoder must retrieve against document embeddings produced by a much stronger offline encoder.
Thus, the problem is not only domain adaptation, but also \emph{embedding-space compatibility}: the compact query encoder must produce representations that are well calibrated to the document space of the teacher retriever.
Direct supervised training of the asymmetric architecture may improve task performance, but it couples multiple objectives like domain adaptation, representation transfer, and online efficiency into a single optimization problem, thus making it challenging to get the high-quality model in one-shot.

\begin{figure*}[t]
    \centering
    \includegraphics[width=0.7\linewidth]{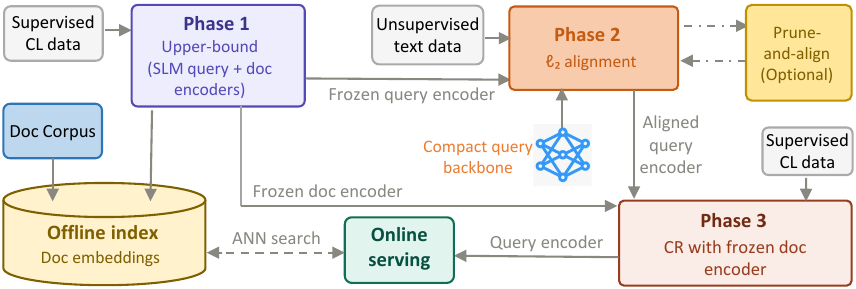}
    \caption{HLM: A three-phase training framework for developing effective and compact SLM retrievers.}
    \Description{HLM: 3-phase training framework}
    \label{fig:harness-lm-overview}
\end{figure*}

At serving time, document representations can be precomputed offline and stored in an ANN index, whereas the query encoder must be executed online for every incoming query. This induces a fundamental asymmetry in model design: the document encoder can afford to be large and computationally intensive, while the query encoder is constrained by strict latency and cost requirements.
In this work, we introduce \textsc{HARNESS}-LM (HLM), a training recipe explicitly designed to exploit this asymmetry. The central philosophy is to decouple offline representation quality from online serving efficiency, enabling high-fidelity document representations without compromising real-time query processing constraints. 

At a high level, the recipe first asks a quality-first question: \emph{how good can an SLM-based retriever become if we relax online deployment constraints?} We answer this by training teacher models that may use larger encoders and/or richer input features. These models are not necessarily deployable, but they define the quality target for the rest of the recipe. Subsequent stages aim to transfer this quality into smaller and more deployable query-side models through alignment, distillation, and latency-aware optimizations.

The HLM training recipe is illustrated in Fig. \ref{fig:harness-lm-overview}. Training proceeds in three phases as follows.
First, we train a high-fidelity reference retriever (referred to as the \emph{teacher}) using larger (4B/8B parameter) SLM encoders and/or richer offline-only features. 
Second, we align a compact query encoder to the teachers' query embedding space using $\ell_2$ regression on unsupervised text data.
Third, we refine the aligned query encoder with supervised Contrastive Learning (CL)~\cite{zhang2025qwen3}, while keeping the teacher document encoder frozen, and refer to it as Contrastive Refinement (CR).
This produces an asymmetric retriever in which the expensive document encoder is used offline for index construction, while the compact query encoder runs online. 

HLM also provides an optional and a practical compression pathway.
After alignment, we progressively prune the compact query encoder by reducing transformer layers and FFN dimensions, followed by re-alignment after each pruning step. This successive prune-and-align strategy is inspired by recent structured model-compression recipes such as the cascaded pruning procedure used in Ministral~3~\cite{liu2026ministral3} and Minitron~\cite{muralidharan2024compact}, but is adapted here to dense retrieval and asymmetric query-document serving.
Unlike unstructured sparsity, our pruning removes entire layers and FFN units, yielding dense smaller models with real latency and cost benefits. In practice, pruning yields up to a $6\times$ latency reduction (40.96 ms $\rightarrow$ 6.8 ms) with only a $1.1\%$  drop in precision (64.3 $\rightarrow$ 63.1), demonstrating a highly favorable quality--efficiency trade-off. 
(refer Table \ref{tab:latency_shortened} for details).

We evaluate HLM on Bing Ads sponsored search retrieval benchmarks and validate the resulting models through online A/B testing on Bing Ads live traffic.
Across offline and online evaluations, HLM model consistently improve retrieval quality over the current ensemble of models in production, with online deployed model showing significant improvements across all key business metrics: $\mathbf{+1\%}$ \textbf{Revenue}, $\mathbf{+0.6\%}$ \textbf{Impression} and $\mathbf{+0.4\%}$ \textbf{Click} uplift without degrading quality.
The main contributions of our work are as follows:
\begin{itemize}[leftmargin=*]
\setlength{\parskip}{0pt}
\setlength{\itemsep}{0pt}
    \item We introduce HLM, a three-phase training recipe for obtaining efficient and high-quality SLMs for sponsored search retrieval. 
    \item We propose $\ell_2$ alignment based optimization that result into effective knowledge transfer from teacher to student.
    \item We propose a progressive prune-and-align strategy to produce encoders based on compute-latency trade-offs.
    \item We conduct an extensive study across multiple design dimensions, including embedding size, feature representations, model architecture, training objectives, and optimization strategies, to develop high-performance SLMs suitable for production settings.
    \item We validate HLM through offline evaluation on Bing Ads retrieval dataset and online A/B testing, showing consistent gains over production baselines while reducing online inference cost.
\end{itemize}

\subsection{Background and Related Work}
\label{sec:related_work}

\textbf{Dense retrieval and CL}:
Dense retrieval uses dual encoders to map queries and documents to a shared embedding space, enabling offline document indexing and low-latency online query encoding. Foundational work like DPR~\cite{karpukhin2020dpr}, ANCE~\cite{xiong2021ance}, and TwinBERT~\cite{lu2020twinbert} established contrastive training with hard negatives and asymmetric serving as key design principles. SamToNe~\cite{moiseev2023samtone} further showed that same-tower negatives improve dual-encoder training. HLM builds on these ideas but focuses on transferring high-capacity SLM retrievers to compact, production-ready retrievers.

\textbf{SLMs as embedding models}:
Recent work shows that decoder-only language models can serve as strong embedding backbones through architectural and training modifications~\cite{ma2024finetuning, behnamghader2024llm2vec, lee2025nvembed, muennighoff2025grit}. Among these, Qwen3-Embedding~\cite{zhang2025qwen3} is particularly relevant as it spans multiple model sizes (0.6B–8B), supports Matryoshka representations, and achieves strong multilingual performance. HLM is a principled approach to specialize such SLM backbones for Ads retrieval while maintaining deployability in a large scale production settings like sponsored search.

\textbf{Compression, pruning, and alignment}:
Deploying SLM-based retrievers under strict latency constraints motivates compression through distillation and pruning. Ministral~3~\cite{liu2026ministral3} uses iterative pruning with continued training to obtain compact models. Kernel-based unsupervised alignment~\cite{pmlr-v267-gong25b} matches representations across encoders without supervised retraining. HLM combines these ideas: we train a high-quality teacher, align a compact query encoder via $\ell_2$ regression, optionally prune, and refine with supervised contrastive learning against the frozen teacher document encoder.

\section{HLM: Training Recipe}

{\bf Notation}:
We consider dense retrieval for sponsored search, where the goal is to retrieve the most relevant documents (Ads) from the document corpus
$\mathcal{D}$ for an incoming query $q$.
We use a dual-encoder retrieval model with a query encoder $f_Q(\cdot)$ and a document encoder $f_D(\cdot)$, which map the query and document texts to normalized $d$-dimensional embedding vectors\footnote{Throughout the paper, we assume that all encoder embeddings have unit $\ell_2$ norm.}.

The relevance score between a query and a document is computed using the inner product:
$s(q,d) = \langle f_Q(q), f_D(d) \rangle.$
Let $f_Q^{T}$ and $f_D^{T}$ denote the teacher query and document encoders, respectively. Let $f_Q^{S}$ denote the smaller student query encoder obtained after alignment with $f_Q^{T}$. 
Next, we describe the first stage of HLM: constructing a high-quality teacher.

\subsection{Phase 1: Teacher}

In the first phase of HLM, we train a teacher which serves two purposes. First, it tells us how much quality is available from stronger SLM backbones and richer
retrieval features. Second, it provides a target for later stages, where we attempt to recover as much of this quality as possible in a compact model.

We obtain teacher models by relaxing one or more production constraints. One axis is model scale: we can increase the number of parameters in the query and document encoders, for example, by using a larger SLM backbone (like Qwen3-Embedding 4B/8B). Since the document encoder is used offline for indexing, a larger document encoder is often easier to justify than a larger query encoder. However, a large query encoder may still be useful for estimating the maximum quality achievable before compression.

A second axis is feature richness. Beyond raw query and document text (deployable features), teacher models may use additional context, such as offline GPT-generated expansions. These enrich semantics and improve offline retrieval, but are often unavailable or too costly at serving time. We term them \emph{oracle features}; models relying on them are treated as oracle or teacher models, not deployable models. 
An example of oracle features for the query ``\texttt{change from pdf into word free}'' is provided in Appendix \ref{app:oracle_feats_eg}.



We train our teacher using the Qwen3-Embedding contrastive objective~\cite{zhang2025qwen3}, which is a modified InfoNCE loss~\cite{oord2018representation}. Unlike vanilla InfoNCE, it enriches the
denominator with mined hard negatives and same-tower query--query/document--document negatives~\cite{moiseev2023samtone},
while masking likely false negatives from the in-batch pool.

\begin{equation}\label{qwen3_loss}
\mathcal{L}_{\mathrm{QwenCL}}(i)
=
-\log
\frac{
\exp(s(q_i,d_i^+)/\tau)
}{
\exp(s(q_i,d_i^+)/\tau)
+
\sum_{u \in \mathcal{N}_i}
\exp(s_u/\tau)
}.
\end{equation}

Here, $\mathcal{N}_i$ contains the valid negatives after false-negative masking: in-batch
query--document negatives, mined hard negatives, and same-tower query--query/document--document
negatives. The score $s_u$ denotes the corresponding similarity term, 
and $\tau$ is the temperature (refer \cite{zhang2025qwen3} for details).

The resulting teacher defines the target for the rest of the HLM recipe. Later phases aim to retain this quality while removing the challenges that make the teacher impractical for production: large latency of
online query-side inference, dependence on unavailable GPT generated features, or excessive serving cost.

\subsection{Phase 2: Alignment}

The teacher obtained in phase-1 provides a high-quality target, but it is not directly suitable for online serving. In particular, the best teachers may use a large query encoder, large document encoder, or training-time
features that are not available in the online path. We therefore next study whether a smaller student query encoder can be made
compatible with the teacher embedding space. 

The goal of alignment is to train the student query encoder: $f_Q^{S}$, such that it can replace the teacher query encoder:$f_Q^{T}$ from phase-1 at serving time while still retrieving against the
document embeddings produced by the encoder: $f_D^{T}$, from phase-1. This gives an asymmetric retriever: the expensive document encoder directly coming from the teacher is used offline for index construction, while the compact query encoder is used online.

\paragraph{Alignment objectives.}
We explored multiple objectives to transfer the teacher query representation to the smaller query encoder.
One class of objectives is score-level or contrastive distillation, inspired by recent embedding model training
recipes such as KaLM-Embedding-V2, where a compact embedding model is improved using contrastive distillation with fine-grained teacher signals \cite{zhao2025kalm}. Another class of objectives aligns pairwise structures rather than individual embeddings. For example, the loss objective for unsupervised kernel alignment~\cite{pmlr-v267-gong25b} aligns representations by matching kernel matrices induced by teacher and student embeddings.

In our setting, we found that a simpler objective works best: direct $\ell_2$ alignment between teacher and student query embeddings. For an alignment corpus $\mathcal{A} = [q_1, q_2, \cdots, q_N]$ consisting of $N$ pieces of texts, we minimize
\begin{equation}\label{eq:l2_loss}
\mathcal{L}_{\text{align}}(f_Q^S) =
\sum_{i=1}^N\|f_Q^{S}(q_i) - f_Q^{T}(q_i)\|_2^2.
\end{equation}
The teacher query encoder is frozen during this stage, and only the student query encoder is updated. After alignment, the student query embedding is scored against the frozen teacher document embedding as 
$s_{\text{align}}(q,d) = \langle f_Q^{S}(q), f_D^{T}(d) \rangle. $



\begin{algorithm}[t]
\caption{Structured Pruning for Student Query Encoder}
\label{alg:structured-pruning}
\begin{algorithmic}[1]
\Require Query encoder $f_Q$ with $L$ transformer layers, calibration corpus $\mathcal{C}$, 
target layer count $K_L$, target FFN dimension $K_F$
\Ensure Pruned query encoder $\widetilde{f}_Q$

\Statex\algcomment{Stage 1: Layer Pruning (Depth Reduction)}
\State Run calibration data through $f_Q$ and collect hidden states at each layer boundary.
\State For each layer $\ell \in \{1, \ldots, L\}$, compute the importance score\footnotemark:
\[
I_{\mathrm{layer}}(\ell)
=
\mathbb{E}_{x \sim \mathcal{C}}
\left[
\|h_{\ell}^{\mathrm{out}}(x)\|_2/\|h_{\ell}^{\mathrm{in}}(x)\|_2
\right]
\]
where $h_{\ell}^{\mathrm{in}}(x), h_{\ell}^{\mathrm{out}}(x) \in \mathbb{R}^d$ are the hidden states 
entering and exiting layer $\ell$, respectively.
\State Retain the top-$K_L$ layers ranked by $I_{\mathrm{layer}}(\ell)$, preserving their original order.

\Statex \algcomment{Stage 2: FFN Pruning (Width Reduction)}
\State For each retained layer $\ell$, score FFN hidden unit $j \in \{1, \ldots, d_{\mathrm{ffn}}\}$:
\[
I_{\mathrm{ffn}}(\ell,j)
=
\mathbb{E}_{x \sim \mathcal{C}}
\left[
\left|
\sigma(W_{\ell}^{g} h_{\ell}(x))_j
\cdot
(W_{\ell}^{u} h_{\ell}(x))_j
\right|
\right]
\]
where $W_{\ell}^{g}, W_{\ell}^{u} \in \mathbb{R}^{d_{\mathrm{ffn}} \times d}$ are the gate and up 
projection matrices of the SwiGLU FFN, and $\sigma(\cdot) = \mathrm{SiLU}(\cdot)$ is the 
Sigmoid Linear Unit activation.
\State For each layer, retain the top-$K_F$ FFN units by $I_{\mathrm{ffn}}(\ell, j)$.
\State Prune corresponding rows from $W_{\ell}^{g}, W_{\ell}^{u}$ and columns from 
$W_{\ell}^{d} \in \mathbb{R}^{d \times d_{\mathrm{ffn}}}$ (down projection).
\State \Return Dense pruned model $\widetilde{f}_Q$ with $K_L$ layers and FFN dimension $K_F$.
\end{algorithmic}
\end{algorithm}

\footnotetext{
The layer importance score $I_{\mathrm{layer}}(\ell)$ measures how much layer $\ell$ 
transforms the representation magnitude. Due to residual connections, 
$h_{\ell}^{\mathrm{out}} = h_{\ell}^{\mathrm{in}} + F_\ell(h_{\ell}^{\mathrm{in}})$ 
where $F_\ell$ is the layer's transformation. A ratio near 1 indicates the layer 
adds little to the representation, making it a candidate for removal. 
}

This objective is attractive for two reasons. First, it directly enforces compatibility between the embedding spaces of the student and teacher query encoders, which is exactly what is required for asymmetric retrieval when the document encoder is frozen. Second, it avoids requiring
labeled query--document pairs during alignment; any large corpus of query-like text can be used to expose the student to the geometry of the teacher query space.

As demonstrated in Sec. \ref{sec:experiments}, when we use Qwen3-Embedding-0.6B as the student query encoder and Qwen3-Embedding-4B as the teacher query encoder, the smaller encoder is able to recover the retrieval quality of the teacher encoder using $\ell_2$ alignment when combined with the frozen teacher document encoder. This result is central to the HLM recipe: it shows that the quality of a large query encoder can be transferred into a smaller encoder without rebuilding the document index with the student model. In other words, alignment lets us retain the stronger offline document encoder while replacing only the online query encoder.

\subsubsection{Alignment under pruning.}
In addition to aligning a smaller encoder to a larger encoder, we use pruning to understand how far the query encoder can be compressed. For pruning, we adopt structured pruning of transformer layers and the feedforward layer within the transformer layers (described in Algorithm \ref{alg:structured-pruning}).  

We adopt successive pruning in the same spirit as the cascade pruning recipe used in Ministral 3~\cite{liu2026ministral3}: rather than pruning directly to the smallest model in one step, we prune successively and re-align after each pruning stage. In our case, we start from the aligned Qwen3-Embedding-0.6B model (which has 28 transformer layers) and progressively prune the model to smaller variants, including 14-layer, 7-layer, 4-layer, and 2-layer models, while also reducing the Feedforward Neural Network (FFN) dimension.

Cascade distillation avoids the knowledge shock of a direct leap by using intermediate models as stepping stones, ensuring the student inherits more stable, pre-refined representations. This allows the "prune-align-repeat" pipeline to ensure that pruning for each smaller model starts from a good aligned checkpoint rather than from a heavily damaged representation \cite{liu2026ministral3,muralidharan2024compact}.
The progressive pruning technique is described in Algorithm \ref{alg:progressive-prune-align}.

\begin{algorithm}[t]
\caption{Progressive prune-and-align pipeline}
\label{alg:progressive-prune-align}
\begin{algorithmic}[1]
\Require Aligned student $f_Q^{S}$, teacher query encoder $f_Q^{T}$, frozen teacher document encoder $f_D^{T}$,
alignment corpus $\mathcal{A}$, pruning targets $\{(K_L^{(r)}, K_F^{(r)})\}_{r=1}^{R}$
\Ensure Compact query encoder $\widehat{f}_Q^{S}$

\State Set $g_Q \leftarrow f_Q^{S}$.
\For{each pruning target $(K_L, K_F)$}
    \State Prune $g_Q$ to $K_L$ layers and FFN dimension $K_F$ using Algorithm~\ref{alg:structured-pruning}.
    \State Re-align $g_Q$ to $f_Q^{T}$ on $\mathcal{A}$ by minimizing $\ell_2$ loss $\mathcal{L}(g_Q)$ from Eq. \ref{eq:l2_loss}.
\EndFor
\State \Return $\widehat{f}_Q^{S} \leftarrow g_Q$
\end{algorithmic}
\end{algorithm}

This progressive prune-and-align strategy gives us a practical way to trace the quality--latency frontier. Larger capacity student models retain more of the teacher quality, while smaller models reduce inference costs (like GPU hardware) and latency. We observed that after successive pruning and re-alignment, the 4-layer pruned encoder is suitable to run on CPUs or low-end GPUs as per the online serving latency requirements at Bing Ads, substantially reducing inference cost compared with GPU-served SLM encoders. While successive pruning provides a systematic pathway to explore progressively smaller architectures, it is not a mandatory component of the alignment pipeline. It can be optionally employed to characterize the quality–latency trade-off frontier, enabling selection of a model variant that best meets target performance and deployment constraints.

\subsection{Phase 3: Contrastive Refinement} 

\begin{algorithm}[t]
\caption{HLM training recipe}
\label{alg:harness-lm}
\begin{algorithmic}[1]
\Require Supervised retrieval data $\mathcal{D}_{\mathrm{sup}}$,
alignment corpus $\mathcal{A}$, large teacher backbone, compact student backbone,
optional pruning targets
\Ensure Deployable asymmetric retriever $(\widehat{f}_Q^{S}, f_D^{T})$
\Statex \algcomment{Phase 1: Train the teacher}
\State Train a teacher $(f_Q^{T}, f_D^{T})$ using large SLM backbone and/or richer offline features with supervised CL loss (Eq. \ref{qwen3_loss}).
\Statex \algcomment{Phase 2: Align compact query encoder}
\State Initialize compact query encoder $f_Q^{S}$ from a pretrained SLM checkpoint.
\State Align $f_Q^{S}$ to the teacher query encoder $f_Q^{T}$ on $\mathcal{A}$ using  $\ell_2$ loss $\mathcal{L}_{\mathrm{align}}(f_Q^S)$ (Eq. \ref{eq:l2_loss})
\State Optionally compress $f_Q^{S}$ using progressive prune-and-align
(Algorithm~\ref{alg:progressive-prune-align}).
\Statex \algcomment{Phase 3: Contrastive refinement}
\State Freeze $f_D^{T}$ and continue supervised contrastive refinement on $f_Q^{S}$ using $\mathcal{D}_{\mathrm{sup}}$ with CL loss.
\State Set $\widehat{f}_Q^{S} \leftarrow f_Q^{S}$.
\State \Return $(\widehat{f}_Q^{S}, f_D^{T})$.
\end{algorithmic}
\end{algorithm}

The alignment phase makes the compact query encoder compatible with the teacher embedding space. However, alignment
alone is not directly optimized for the downstream retrieval objective. It teaches the student query encoder to mimic
the teacher query representation, but it does not explicitly train the student to separate positive documents from hard
negatives under the final asymmetric retrieval setup. We therefore add a final stage of refinement using contrastive loss function and refer to it as Contrastive Refinement (CR).

Starting from the aligned student query encoder $f_Q^{S}$, we freeze the teacher document encoder $f_D^{T}$ and continue training only the student query encoder using supervised query--document pairs. 
We optimize the Qwen3 CL loss from Eq. \ref{qwen3_loss} for CR phase.
The document encoder remains frozen throughout this stage, so the online query encoder is adapted to the fixed document representation space. The HLM training recipe is illustrated in Fig. \ref{fig:harness-lm-overview} and described in Algorithm \ref{alg:harness-lm}.

Empirically (as noted in Sec. \ref{sec:experiments}), this final contrastive stage improves precision by a few points over the aligned checkpoint. 
This is because the aligned encoder $f_Q^{S}$ provides a good optimization starting point by already being in the teacher-compatible embedding space. Then, the contrastive objective can focus on local refinements: increasing the margin between positives and hard negatives, correcting task-specific errors, and improving top-$K$ precision. 

Additionally, we also note in Sec. \ref{sec:experiments} that the three-stage HLM recipe significantly outperforms one-shot asymmetric supervised CL model with the smaller query and larger document encoders. 
We hypothesize that this happens for the following reasons. 

First, HLM separates three coupled objectives: learning a strong document space, transferring the compact query encoder into that space, and then adapting it for sponsored-search retrieval. In contrast, one-shot asymmetric model expect the compact query encoder to learn both task discrimination and cross-encoder compatibility from only pairwise positive/negative labels; this supervision from sparse information is better used after the student has absorbed task-relevant structure from the larger model through alignment.



Second, HLM prevents the document space from being constraint by the capacity-limited query encoder. When both encoders are trained together, the larger document encoder can co-adapt to representations that are convenient for the smaller query encoder, rather than striving for the high quality document representations that it can possibly learn. By freezing the teacher document encoder, HLM keeps the offline index anchored in the best available document space and forces the compact query encoder to become compatible with it.



\section{Experiments \& Results}
\label{sec:experiments}

{\bf Experimental Setup}:
Our supervised training data consists of 250M query--document pairs from Bing Ads (used in Phases 1 and 3). For phase-2, we use 2B query texts as the dataset for $\ell_2$ alignment (note that this is unlabeled data, and hence, easier to curate). All our datasets are global, multilingual and obtained from different query deciles\footnote{Queries are classified into head, torso and tail deciles based on their Search Results page views}. Unless stated otherwise, we use an embedding dimension of $d=128$ and one hard negative per query--document pair for contrastive learning. Hard negatives are mined using a high-quality relevance model, where documents with scores in $(0.2, 0.5]$ are treated as hard negatives and those with scores $>0.5$ as positives. 

We train phase-1 (teacher) model on 16 NVIDIA A100 80GB GPUs, and Phases 2 and 3 on 16 NVIDIA A100 40GB GPUs. Training uses a linearly warmed-up learning rate from zero over the first 10\% of data, followed by linear decay, and the peak learning rate is selected via hyperparameter search. Optimization uses Adam \cite{kingma2014adam} and 
batch sizes are chosen to maximize GPU utilization. 
Unless stated otherwise, we prepend the following prompt to queries during phase-1: \texttt{"Given a web search query, retrieve relevant passages that answer the query.\textbackslash n \{user query:\}"}.

We evaluate HLM on an internal Bing Ads retrieval benchmark sampled from real sponsored search data. The evaluation set contains 230K user queries from 160+ countries and 50+ languages, while the retrieval corpus contains 47 million Ads. Each model encodes queries and Ads into dense vectors; ad embeddings are stored in a DiskANN vector database \cite{jayaram2019diskann}, and the top-100 Ads are selected by inner-product similarity with the query embedding. To measure retrieval quality, we score the retrieved query--ad pairs with a high-quality relevance model and report Precision@100 (P@100), defined as the fraction of top-100 retrieved Ads predicted to be relevant.

We use the Qwen3 family of embedding models \cite{zhang2025qwen3} as the backbone for all experiments for the following reasons: 1) it spans multiple capacities: 0.6B, 4B, and 8B, 2) achieves strong performance on public embedding benchmarks such as MMTEB, and 3) supports Matryoshka Representation Learning (MRL) \cite{NEURIPS2022_c32319f4}, enabling flexible performance-capacity tradeoffs through embedding truncation. These properties allow us to study both the quality gains from larger encoders and the constraints of compact online query encoders within a single model family.

\subsection{Teacher}

We first establish quality upper bounds by training teachers using symmetric dual encoders (that is, both query and document encoders have the same architecture but no shared parameters) and comparing their performance across different backbone sizes.


To understand what contributes to the upper bound, we run controlled ablations over several key design choices: model parameters, embedding dimension ($d$), optimization function (traditional InfoNCE vs. QwenInfoNCE loss), prompt in query input (that is, with vs. without prompts), feature set (deployable vs. oracle input features), number of mined negatives, and the number of encoders (shared common encoder vs. dual-encoder architecture for query and documents).

{\bf Model Parameters}:
In Table~\ref{tab:upperbound-scale}, we study the effect of model scale by training symmetric retrievers with Qwen3-Embedding backbones of size 0.6B, 4B, and 8B for both query and document encoders. We adopt two different fine-tuning settings: Full Fine-Tuning (FFT) and Low-Rank Adaptation (LoRA)~\cite{lora}.

For the 4B backbone, increasing the LoRA rank from 16 to 128 improves P@100 from 60.2 to 62.4, as higher rank enables a larger effective parameter budget during fine-tuning. Interestingly, FFT underperforms LoRA (56.4 vs 62.4) for Qwen-4B, which we attribute to two factors: (i) overfitting due to the large number of trainable parameters relative to task complexity, and (ii) reduced batch size (32$\rightarrow$8) due to memory constraints, leading to fewer in-batch negatives during training which can degrade CL performance \cite{chen2020simclr}.

To further increase capacity, we scale both the backbone and LoRA rank, training an 8B--8B model with rank 256, which achieves the best performance of 64.8 P@100\footnote{We limit our exploration of upper bounds to a small set of FFT and LoRA variants, as Phase~1 training is compute-intensive and constrained by GPU availability.}. 
The best performing Qwen-0.6B-0.6B model is 4-6 points worse than its 4B/8B counterparts.
Overall, these experiments establish strong upper bounds at two operating points: 4B--4B (62.4 P@100) and 8B--8B (64.8 P@100). 

\begin{table}[t]
\centering
 
\setlength{\tabcolsep}{3pt}
\caption{Teacher retrieval performance for symmetric Qwen3 retrievers when varying parameters and finetuning strategy.}
\label{tab:upperbound-scale}
\begin{tabularx}{\linewidth}{Y Y c c}
\toprule
\textbf{Model} & \textbf{Setting} & \textbf{P@100} \\
\midrule
Qwen3-0.6B--0.6B & FFT & 58.1 \\
Qwen3-0.6B--0.6B & LoRA (r=64) & 47.6 \\
Qwen3-0.6B--0.6B & LoRA (r=128) & 48.2 \\ 
\midrule
Qwen3-4B--4B & LoRA (r=16) & 60.2 \\
Qwen3-4B--4B & LoRA (r=32) & 61.5 \\
Qwen3-4B--4B & LoRA (r=128) & \textbf{62.4} \\
Qwen3-4B--4B & FFT & 56.4 \\
\midrule
Qwen3-8B--8B & LoRA (r=256) & \textbf{64.8} \\
\bottomrule
\end{tabularx}
\end{table}

{\bf Embedding dimension}:
In Table \ref{tab:upperbound-dim}, we study how much of the teacher retrieval quality depends on the embedding dimension $d$. Smaller embeddings reduce index storage and retrieval cost, but may also limit the capacity of the retrieval space. Instead of retraining models for different $d$, we leverage MRL to obtain lower-dimensional embeddings via truncation. For reference, the maximum embedding dimensions for Qwen3-4B and Qwen3-8B are 2560 and 4096, respectively. We also consider a zero-shot setting, where the open-source model is used without any fine-tuning on proprietary data.

\begin{table}[t]
\centering
\setlength{\tabcolsep}{4pt}
\caption{Effect of embedding dimension. Larger dimensions yield bigger gains for zero-shot models, while improvements saturate for fine-tuned models.}
\label{tab:upperbound-dim}
\begin{tabularx}{\linewidth}{Y c c c}
\toprule
\textbf{Model} & \textbf{Setting} & \textbf{Dim. ($d$)} & \textbf{P@100} \\
\midrule
Qwen-4B--4B & Zero-shot & 128 & 36.1 \\
Qwen-4B--4B & Zero-shot & 2560 & 49.3 \\
\midrule
Qwen-8B--8B & Zero-shot & 128 & 38.5 \\
Qwen-8B--8B & Zero-shot & 2048 & 49.5 \\
Qwen-8B--8B & Zero-shot & 4096 & 50.0 \\
\midrule
Qwen-4B--4B & LoRA (r=128) & 128 & 62.4 \\
Qwen-4B--4B & LoRA (r=128) & 2560 & 66.8 \\
\midrule
Qwen-8B--8B & LoRA (r=256) & 128 & 64.8 \\
Qwen-8B--8B & LoRA (r=256) & 2048 & 67.7 \\
Qwen-8B--8B & LoRA (r=256) & 4096 & 67.8 \\
\bottomrule
\end{tabularx}
\end{table}

We observe two clear trends. First, the gap between small and large $d$ shrinks significantly after fine-tuning (e.g., +13.2 for zero-shot versus +4.4 for LoRA in 4B), suggesting that task-specific adaptation compensates for reduced embedding capacity. Second, for fine-tuned models, performance largely saturates beyond $d=2048$, with negligible improvements up to $d=4096$, indicating that moderate embedding sizes are sufficient for this task.

{\bf Number of hard negatives}:
We study the impact of increasing the number of hard negatives (HN) per query for a Qwen3-4B--4B model fine-tuned with LoRA (rank=128). Moving from no hard negatives to one yields a large gain (59.2 $\rightarrow$ 62.4 P@100), highlighting the importance of informative negatives for contrastive learning. However, increasing from one to five hard negatives provides negligible improvement (62.4 $\rightarrow$ 62.5). We attribute this to two factors: (i) for strong pretrained models, a small number of informative hard negatives is often sufficient, and (ii) adding more hard negatives reduces the effective batch size (32$\rightarrow$12), leading to fewer in-batch negatives and a less favorable training regime \cite{chen2020simclr}.

For the next set of ablations, we fix the backbone to Qwen3-0.6B--0.6B to reduce computational cost. Table~\ref{tab:upperbound-ablations} summarizes results across three axes: prompt usage, feature richness, and loss function.

\begin{table}[t]
\centering
\setlength{\tabcolsep}{4pt}
\caption{Ablations on prompt, features, and loss.}
\label{tab:upperbound-ablations}
\begin{tabularx}{\linewidth}{Y Y c}
\toprule
\textbf{Ablation} & \textbf{Setting} & \textbf{P@100} \\
\midrule
\textit{Prompt in query} & Zero-shot, no prompt & 29.5 \\
& Zero-shot, with prompt & 31.9 \\
& FFT, no prompt & 55.6 \\
& FFT, with prompt & 58.1 \\
\midrule
\textit{Feature set} & Deployable features & 58.1 \\
 & Oracle features & 63.7 \\
\midrule
\textit{Loss function} & Vanilla InfoNCE & 45.1 \\
& Qwen3 objective & 58.1 \\
\bottomrule
\end{tabularx}
\end{table}

{\bf Prompt in query:}  We evaluate the impact of adding a retrieval prompt to the query. Such prompts can help the model better interpret the input as a retrieval task, but they also increase sequence length and hence online inference cost. Prompting consistently improves retrieval quality in both zero-shot (+2.4) and fine-tuned (+2.5) settings, suggesting that prompts act as an effective inductive bias, though at the cost of increased inference overhead.

{\bf Oracle features:} 
In addition to scaling model capacity, we study an orthogonal axis, feature richness. Specifically, we consider the oracle teacher models that augment raw query and document text with additional offline-generated context (e.g., GPT-based query/document expansions) that are not always available in the online serving path 
(see Appendix \ref{app:oracle_feats_eg} for an example).
Oracle features yield a +5.6 P@100 gain, showing that richer context can provide a strong teacher even with a small backbone---and that improvements can come from better features, not just larger models. In Sec.~\ref{exps:alignment}, we explain distilling these benefits into a production-feasible student.

{\bf Loss function:} We also study the impact of the contrastive loss, where negative construction plays a critical role. We compare vanilla InfoNCE~\cite{oord2018representation}, which relies on in-batch negatives, with the Qwen3 contrastive objective~\cite{zhang2025qwen3}, which augments training with same-tower negatives~\cite{moiseev2023samtone}, in-batch negatives, and hard-negatives along with masking false negatives. Using Qwen3 objective leads to a substantial improvement (+13.0 P@100), highlighting the importance of richer negative construction. 
Thus, we adopt it as the default CL objective in HLM, since it provides a stronger and more targeted training signal for retrieval than vanilla InfoNCE.

Overall, our teacher analysis shows that larger backbones, richer features, and improved contrastive objectives all contribute to retrieval quality. In the $d{=}128$ setting,\footnote{We use $d{=}128$ for deployment due to its favorable performance--latency trade-off.} the best teachers achieve 64.8 P@100 (Qwen3-8B), 62.4 (Qwen3-4B), and 63.7 (Qwen3-0.6B Oracle), which we use for alignment and compression, as described next.

\subsection{Query Alignment}
\label{exps:alignment}

The teacher provides a high-quality embedding space, but its query encoder is too expensive for online serving (see Table \ref{tab:latency_shortened} for details). 
Next, we study whether a compact query encoder can be aligned to the teacher query encoder while retrieving against the frozen teacher document encoder.

{\bf Alignment to different upper bounds}:
We first study how the choice of teacher affects alignment quality. Table~\ref{tab:align-teacher} shows both the teacher performance and the aligned student performance, allowing us to measure the \emph{transfer gap}. Stronger teachers produce stronger students, but the gap also grows  (0.0 $\rightarrow$ 2.3), suggesting that transferring a more powerful retrieval space into a compact student becomes progressively harder.

\begin{table}[t]
\centering
\setlength{\tabcolsep}{4pt}
\caption{Effect of teacher quality on query encoder alignment with student backbone as Qwen3-0.6B}
\label{tab:align-teacher}
\begin{tabularx}{\linewidth}{Y c c c}
\toprule
\textbf{Teacher} & \textbf{Teacher} & \textbf{Student} & \textbf{Gap} \\
& \textbf{(P@100)} & \textbf{(P@100)} & \\
\midrule
Qwen3-4B zero-shot & 36.1 & 36.1 & 0.0 \\
Qwen3-4B LoRA (r=32) & 61.5 & 61.2 & 0.3 \\
Qwen3-4B LoRA (r=128) & 62.4 & 61.9 & 0.5 \\
Qwen3-8B LoRA (r=256) & 64.8 & 62.5 & 2.3 \\
\bottomrule
\end{tabularx}
\end{table}


{\bf Alignment to oracle teachers:}
We align the student to an oracle teacher using richer, non-deployable features while restricting student inputs to deployable features only. The oracle teacher achieves 63.7 P@100; the aligned student reaches 61.7, leaving a gap of 2.0---comparable to the 8B teacher (gap of 2.3). This indicates that stronger teachers yield larger transfer gaps, and that alignment can distill some, but not all, benefit of richer offline context.

For the remaining ablations, we fix the teacher to Qwen3-4B LoRA (rank=128), with teacher P@100 of 62.4, and the student to Qwen3-0.6B. Table~\ref{tab:align-ablations} summarizes the results.

\begin{table}[t]
\centering
\caption{Ablations for query alignment with fixed teacher Qwen3-4B LoRA (rank=128).}
\label{tab:align-ablations}
\begin{tabularx}{\linewidth}{Y c c c}
\toprule
\textbf{Setting} & \textbf{Teacher} & \textbf{Student} & \textbf{Gap} \\
\midrule
Without prompt & 62.4 & 61.9 & 0.5 \\
With prompt & 62.4 & 61.9 & 0.5 \\
\midrule
Ads queries & 62.4 & 61.9 & 0.5 \\
Public text + Ads queries & 62.4 & 61.8 & 0.6 \\
\midrule
Pretrained Qwen3-0.6B & 62.4 & 61.9 & 0.5 \\
Random initialization & 62.4 & 59.7 & 2.7 \\
\midrule
KL divergence \cite{zhao2025kalm} & 62.4 & 56.4 & 5.8 \\
Kernel matrix alignment \cite{pmlr-v267-gong25b} & 62.4 & 58.7 & 3.7 \\
\bottomrule
\end{tabularx}
\end{table}

{\bf Prompt in Query}: Unlike supervised contrastive training, prompt usage provides no benefit during alignment. We therefore omit the prompt from the student model's input but keep it in the teacher model during the alignment phase in HLM, which simplifies serving and avoids additional online latency.

{\bf Public data}: Alignment on Ads queries alone preserves task-specific retrieval quality, but substantially degrades general embedding quality. The aligned model trained only on Ads queries achieves 23.0 NDCG@10 on MMTEB, far below the 4B--4B upper bound of 42.1. Adding public text from \cite{oepen2025hplt}, covering over 200 languages, in a 50-50 mixture with Ads queries leaves Ads performance nearly unchanged (61.8 vs.\ 61.9), while improving MMTEB from 23.0 to 39.8. This shows that broader alignment data preserves general embedding quality without sacrificing Ads-task performance.


\begin{figure}[t]
    \centering
    \includegraphics[width=0.92\linewidth]{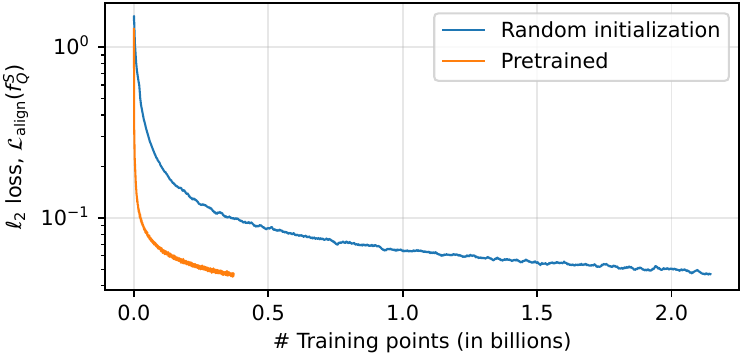}
    \caption{Alignment loss (Eq. \ref{eq:l2_loss}) as a function of training points for a randomly initialized model versus a pretrained model, showing that pretraining yields significantly faster convergence and a lower final loss.}
    \label{fig:random_init_comp}
    \Description{Alignment loss}
\end{figure}

{\bf Pretraining}: Pretraining is critical for both alignment quality and data efficiency. Compared to a pretrained (zero-shot) student, randomly initialized student performs worse (59.7 vs. 61.9 P@100) and requires substantially more training data (2.2B vs. 0.4B), as shown in Fig.~\ref{fig:random_init_comp}. This indicates that the pretrained checkpoint starts from a more favorable optimization region for alignment.



{\bf Alignment loss function}: One of the key contribution of our paper is the proposed $\ell_2$ alignment objective in Eq.~\ref{eq:l2_loss}. Although simple to implement and fully unsupervised, it outperforms more elaborate alternatives such as Kullback Leibler (KL) divergence based contrastive distillation with teacher soft signals~\cite{zhao2025kalm} and kernel-matrix alignment of teacher and student similarities~\cite{pmlr-v267-gong25b}, as also shown in Table~\ref{tab:align-ablations}. We believe this is because direct embedding-level supervision provides a denser and more faithful transfer signal than matching softened scores or pairwise structure alone, while naturally preserving compatibility with the frozen document encoder. 
These loss functions are described in detail in Appendix \ref{app:other_align_loss_fn}.

\subsection{Progressive pruning}

We next study whether the aligned Qwen3-0.6B query encoder can be compressed for a better performance-latency tradeoff. We adopt structured pruning of self-attention and feedforward layers, followed by re-alignment to the frozen Qwen3-4B LoRA (rank=128) teacher. 

\begin{table}[t]
\centering
\setlength{\tabcolsep}{4pt}
\caption{Pruning of the aligned Qwen3-0.6B query encoder.}
\label{tab:pruning}
\begin{tabularx}{\linewidth}{Y c c c}
\toprule
\textbf{Student model} & \textbf{Layers} & \textbf{FFN dim} & \textbf{P@100} \\
\midrule
Aligned Qwen3-0.6B & 28 & 3072 & 61.9 \\
Pruned + aligned & 14 & 2304 & 61.6 \\
Pruned + aligned & 7 & 1536 & 61.0 \\
Pruned + aligned & 4 & 1024 & 60.4 \\
Pruned + aligned & 2 & 1024 & 56.5 \\
\midrule
Pruned directly + aligned & 4 & 1024 & 50.2 \\
\bottomrule
\end{tabularx}
\end{table}

Table \ref{tab:pruning} shows the impact of progressing pruning and alignment. We observe a smooth degradation as the model is progressively pruned, with only modest loss up to 4 layers (61.9 $\rightarrow$ 60.4), suggesting that much of the retrieval behavior can be preserved in substantially smaller students for our task. In contrast, the 2-layer model shows a sharper drop (56.5), indicating that capacity becomes the main bottleneck at extreme compression.

A key observation is that pruning must be done progressively. When we prune the zero-shot Qwen3-Embedding-0.6B model directly to 4 layers and then align, performance collapses to 50.2 even after training on 2B points. This mirrors our earlier pretraining result from Fig. \ref{fig:random_init_comp}: alignment works best when the student starts from a good pretrained checkpoint that remains close to the original model, and becomes much harder when optimization begins from a severely compressed or poorly initialized model. 

\subsection{Contrastive refinement}
We next evaluate the final phase of HLM, i.e, Contrastive Refinement (CR). Starting from the aligned student query encoder from phase-2, we freeze the teacher document encoder and continue training only the query encoder with the Qwen3 contrastive loss (Eq. \ref{qwen3_loss}). We do not prepend a retrieval prompt during this phase as the aligned model did not use the query prompt. We consider two teacher document towers: Qwen3-4B LoRA (rank=128), with P@100 of 62.4, and Qwen3-8B LoRA (rank=256), with P@100 of 64.8.

\begin{table}[t]
\centering
\setlength{\tabcolsep}{4pt}
\caption{Effect of Contrastive Refinement (CR) on P@100.}
\label{tab:cr}
\begin{tabularx}{\linewidth}{Y Y c c c}
\toprule
\textbf{Teacher (frozen)} & \textbf{Student} & \textbf{Before CR} & \textbf{After CR} & \textbf{Gain} \\
\midrule
Qwen3-4B LoRA (r=128) & Aligned Qwen3-0.6B & 61.9 & 64.2 & +2.3 \\
Qwen3-4B LoRA (r=128) & Pruned \& Aligned 4L & 60.4 & 63.1 & +2.7 \\
Qwen3-8B LoRA (r=256) & Aligned Qwen3-0.6B & 62.5 & 64.3 & +1.8 \\
\bottomrule
\end{tabularx}
\end{table}

\begin{figure*}[t]
    \centering
    \begin{subfigure}[b]{0.23\linewidth}
        \centering
        \includegraphics[width=\linewidth]{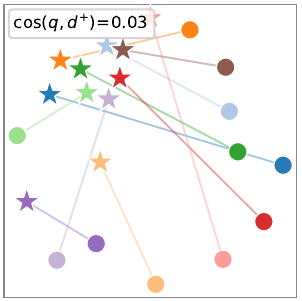}
        \caption{Zero-Shot}
        \label{fig:emb_zero_shot}
    \end{subfigure}
    \hspace{2em}
    \begin{subfigure}[b]{0.23\linewidth}
        \centering
        \includegraphics[width=\linewidth]{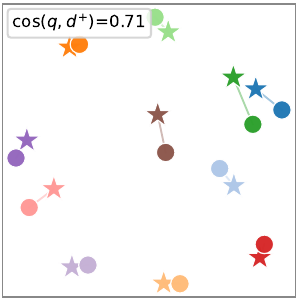}
        \caption{Phase 2: Aligned}
        \label{fig:emb_aligned}
    \end{subfigure}
    \hspace{2em}
    \begin{subfigure}[b]{0.23\linewidth}
        \centering
        \includegraphics[width=\linewidth]{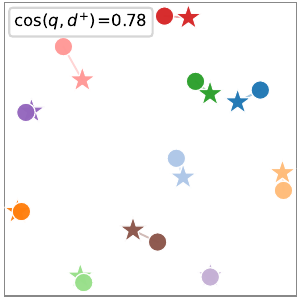}
        \caption{Phase 3: CR}
        \label{fig:emb_cr}
    \end{subfigure}
    
    \caption{2-D projection of query (stars) and document (circles) embeddings across HLM training phases.}
    \label{fig:emb_visual}
    \Description{2-D embedding visualization}
\end{figure*}

Table \ref{tab:cr} shows the impact of CR on the aligned models. CR consistently improves retrieval quality over the aligned checkpoints. With the 4B teacher, refinement improves the aligned Qwen3-0.6B student from 61.9 to 64.2, and even the progressively pruned 4-layer student from 60.4 to 63.1. With the stronger 8B teacher, refinement improves the aligned Qwen3-0.6B student from 62.5 to 64.3. These results support our hypothesis that alignment provides a strong initialization in the teacher-compatible embedding space, after which contrastive training can focus on local ranking refinements like increasing margins between positives and hard negatives and improving compatibility with the frozen document index.

An interesting finding is that the gain from CR is larger for the pruned 4-layer student than for the full 0.6B student (+2.7 vs.\ +2.3 with the 4B teacher). This suggests that although pruning introduces a larger gap after alignment, much of that gap is recoverable through task-specific refinement. In other words, alignment transfers compatibility, while CR recovers discrimination.


Fig~\ref{fig:emb_visual} shows a Multi-Dimensional Scaling (MDS) projection of 
cosine distances at different phases in HLM. In the zero-shot setting (Fig. \ref{fig:emb_zero_shot}), the Qwen3-0.6B query 
encoder produces embeddings that lie farther away from the document embeddings, 
yielding near-zero similarity between matched pairs ($\cos(q,d^{+})\!=\!0.03$), where $d^+$ is the right document for query $q$. 
After $L_2$ alignment in phase-2 (Fig. \ref{fig:emb_aligned}), query embeddings shift toward their corresponding 
documents, raising average similarity to $0.71$. Contrastive refinement in phase-3 (Fig. \ref{fig:emb_cr}) further 
tightens matched pairs while separating unrelated ones, achieving $\cos(q,d^{+})\!=\!0.78$, averaged across all $(q,d^+)$ pairs.

\begin{table}[t]
\centering
\setlength{\tabcolsep}{3pt}
\caption{HLM vs. asymmetric fine-tuning.}
\label{tab:cr-vs-fft}
\begin{tabularx}{\linewidth}{c c c | c}
\toprule
\textbf{Query} & \textbf{Document} & \textbf{Naive CL} & \textbf{HLM} \\
\midrule
Qwen3-0.6B (FFT) & Qwen3-4B (FFT) & 53.4 & \multirow{2}{*}{64.2} \\
Qwen3-0.6B (FFT) & Qwen3-4B (LoRA, r=128) & 54.6 &  \\
\midrule
Qwen3-0.6B (FFT) & Qwen3-8B (LoRA, r=256) & 56.2 & 64.3 \\
\bottomrule
\end{tabularx}
\end{table}

\begin{table*}[t]
\centering
\setlength{\tabcolsep}{4pt}
\caption{Latency–quality trade-offs of HLM query encoders. Latency and throughput are measured on a Nvidia A100 GPU.}
\label{tab:latency_shortened}
\begin{tabularx}{\linewidth}{Y c c c c c c c c}
\toprule
\textbf{Query} & \textbf{Hidden} & \textbf{FFN} & \textbf{Query} & \textbf{Key-Value} & \textbf{Number of} & \textbf{Latency} & \textbf{Throughput} & \textbf{HLM} \\
\textbf{Encoder} & \textbf{dim} & \textbf{dim} & \textbf{Heads} & \textbf{Heads} & \textbf{Parameters} & \textbf{(ms)} & \textbf{(queries/sec)} & \textbf{P@100} \\
\midrule
Qwen3-8B (CL) & 4096 & 12288 & 32 & 8 & 7.57B & 186.14 & 338 & 64.8 \\
Qwen3-4B (CL) & 2560 & 9728 & 32 & 8 & 4.02B & 117.81 & 530 & 62.4 \\
Qwen3-0.6B (Align + CR) & 1024 & 3072 & 16 & 8 & 0.60B & 40.96 & 1,461 & 64.3 \\
Pruned 4L layers (Align + CR) & 1024 & 1024 & 16 & 8 & 0.19B & 6.80 & 6,808 & 63.1 \\
\bottomrule
\end{tabularx}
\end{table*}

{\bf One-shot asymmetric fine-tuning}:
Next, we compare HLM with the standard alternative of directly fine-tuning an asymmetric dual-encoder retriever with a small query and a larger document encoder.
As shown in Table \ref{tab:cr-vs-fft}, HLM substantially outperforms one-shot asymmetric fine-tuning, with gains of 8.1-9.6 P@100 over the direct FFT baselines. This confirms the benefit of the three-phase recipe: direct asymmetric training must learn cross-tower compatibility and task discrimination jointly, whereas HLM first transfers the compact query encoder into the teacher-compatible space and then applies task-specific refinement.

{\bf Latency vs. Performance}: Table~\ref{tab:latency_shortened} presents a detailed comparison of the latency--quality trade-off across query encoders obtained at different stages of the HLM recipe. As expected, larger models such as Qwen3-8B and Qwen3-4B achieve strong retrieval performance but incur significantly higher latency (186.1 ms and 117.8 ms, respectively). In contrast, the aligned and contrastively refined Qwen3-0.6B model matches the 8B model's performance (64.2 vs. 64.8 P@100) while reducing latency by more than $4\times$ and increasing throughput from 338 to 1,461 queries/sec. Further compression via structured pruning yields a 4-layer model with only 0.19B parameters that achieves comparable performance (63.1 P@100) at just 6.8 ms latency and over 6,800 queries/sec throughput. These results highlight that HLM effectively preserves retrieval quality while enabling substantial gains in efficiency through alignment and progressive compression.

\begin{table}[t]
\centering
\caption{Online A/B results of HLM on sponsored search traffic (relative change with respect to production)}
\label{tab:online_ab}
\setlength{\tabcolsep}{6pt}
\begin{tabular}{ccccc}
\toprule
\textbf{Revenue} & \textbf{Clicks} & \textbf{Impressions} & \textbf{QBR} & \textbf{Ad Defect} \\
\midrule
+1.0\% & +0.4\% & +0.38\% & $\sim$0\% & $\sim$0\% \\
\bottomrule
\end{tabular}
\end{table}

\subsection{Online A/B Testing}
To evaluate the effectiveness of the HLM training recipe in a real-world setting, we conducted large-scale online A/B experiments on live traffic from Bing sponsored search. We choose the pruned 4-layer model with 190 million parameters for deployment due to its good latency-quality trade-off. Table~\ref{tab:latency_shortened} for more details on the latency--quality trade-offs and architectural configurations.

The HLM-based retriever was deployed in production and benchmarked against an ensemble of state-of-the-art retrieval approaches already running in production.
We report key business and quality metrics, including revenue, Ad clicks, Ad impressions, Quick Back Rate (QBR), and Ad defect rate. QBR measures the fraction of ad clicks that result in users quickly returning back from the clicked Ad, serving as a proxy for poor user satisfaction. Ad Defect rate, computed using high-quality offline relevance models, captures the proportion of irrelevant Ads shown.

From Table~\ref{tab:online_ab}, we see that HLM achieves consistent gains across all primary engagement and revenue metrics, while maintaining QBR and Ad defect rates at parity with the baseline. These results demonstrate that the HLM recipe enables deployable SLMs that improve both user engagement and monetization without compromising quality, all within strict production latency constraints.

\section{Conclusion}


We present \textsc{HARNESS}-LM, a three-phase training recipe that decouples representation transfer from task-specific optimization. Extensive ablations demonstrate that this decoupling is crucial, with direct asymmetric fine-tuning underperforming by over 10 absolute P@100 points. Our iterative prune-and-align procedure, followed by contrastive refinement, recovers most of the quality lost during compression (P@100 $64.3 \rightarrow 63.1$ for a 4-layer model). The resulting 190M parameter model achieves $27\times$ lower latency while incurring only a 1.7 P@100 drop relative to an 8B teacher. Large-scale online A/B experiments on real-world Bing Ads sponsored search traffic further validate its effectiveness, yielding consistent gains of $\mathbf{+1\%}$ \textbf{in revenue}, $\mathbf{+0.6\%}$ \textbf{in impression}, and $\mathbf{+0.4\%}$ \textbf{in clicks} over the current production ensemble. Beyond empirical gains, our study provides practical insights across key design dimensions for effectively harnessing SLMs in real-world retrieval systems. Scaling HLM to stronger teachers and extending it to broader embedding-based tasks beyond retrieval remain important directions for future work.

\begin{acks}
We thank Anshul Mittal for his help with training-data curation and index creation. We also thank Rakshith V. for his help with curating the public dataset for alignment and for evaluating the HLM models on MMTEB.
\end{acks}


\clearpage
\bibliographystyle{unsrtnat}
\bibliography{references}

\clearpage
\appendix
\section{Oracle Features: Example}
\label{app:oracle_feats_eg}
For the query ``\texttt{change from pdf into word free}'', the corresponding oracle input augments the raw query with multiple feature-rich fields, including query rewrites, an intent explanation \footnote{query rewrites and intent explanations are generated via LLM}, and retrieved web-search titles. Table \ref{tab:oracle-input-example} highlights different features that augments the query. The query, its several rewrites, intent and web-search titles are concatenated and fed to the oracle model while training. 

\begin{table}[h]
\centering
\setlength{\tabcolsep}{3pt}
\renewcommand{\arraystretch}{1.08}
\caption{Example of an oracle input that augments the raw query with feature-rich fields.}
\label{tab:oracle-input-example}
\begin{tabularx}{\linewidth}{@{}p{0.22\linewidth}X@{}}
\toprule
\textbf{Field} & \textbf{Content} \\
\midrule
Query &
\texttt{change from pdf into word free} \\
\addlinespace[2pt]

Rewrites &
\begin{itemize}[leftmargin=*, nosep]
  \item change pdf format to word for free
  \item convert pdf files to word documents free
  \item convert pdf to word online free
  \item free online tool to convert pdf to word
  \item free pdf to word converter
  \item free pdf to word online
  \item how to change pdf to word for free
  \item online pdf to word conversion free
  \item pdf to word converter free download
  \item pdf to word free software
\end{itemize} \\
\addlinespace[2pt]

Query Intent Explanation &
The user is looking for a product, service, or information that can help convert a PDF file into a Word document for free. \\
\addlinespace[2pt]

Web Search Titles &
\begin{itemize}[leftmargin=*, nosep]
  \item convert pdf to word online for free adobe acrobat
  \item pdf to word converter 100 free
  \item pdf to word convert pdf to word online for free
  \item PDF to WORD Convert PDF to Word online for free iLovePDF
\end{itemize} \\
\bottomrule
\end{tabularx}
\end{table}



\section{Alternative Loss Functions for Alignment}
\label{app:other_align_loss_fn}

In this section, we describe the two alternative
alignment objectives we benchmark against the proposed $\ell_2$ loss in
Table~\ref{tab:align-ablations}: the KL-based contrastive distillation loss of
KaLM-Embedding-V2~\cite{zhao2025kalm} and the Kernel-based Unsupervised Embedding
Alignment (KUEA) loss of Gong et al.~\cite{pmlr-v267-gong25b}.

Let $f_Q^{T}$ and $f_D^{T}$ be the teacher query and document encoders, respectively,  from the upper-bound model of Phase~1. In our experimental setting, this is the Qwen3-4B model trained with LoRA rank 128. Let
$f_Q^{S}$ be the student query encoder (Qwen3-0.6B) that we are
aligning to the 4B-query encoder. Also, a reminder that during the alignment phase, both $f_Q^{T}$ and $f_D^{T}$ remain frozen.

\subsection{KL-based contrastive distillation}
\label{app:kalm_loss}

In the Kullback-Leibler divergence-based loss function defined in~\cite{zhao2025kalm}, the loss function transfers the teacher's \emph{score
distribution} over a candidate set to the student. Concretely, for each query
$q_i$ in a mini-batch $\{(q_i, d_i^{+})\}_{i=1}^{B}$, it forms a candidate set
$\mathcal{C}_i$ consisting of its positive $d_i^{+}$ together with the other
in-batch positives $\{d_j^{+}\}_{j \neq i}$ as hard negatives. Because the
document encoder is frozen, the candidate embeddings
$\{f_D^{T}(d)\}_{d \in \mathcal{C}_i}$ are produced once by the teacher and
reused for both the teacher and the student forward pass.
The teacher and student induce two similarity distributions over
$\mathcal{C}_i$,
\[
p_i^{T}(d) \;=\; \frac{\exp\!\big(f_Q^{T}(q_i)^{\top} f_D^{T}(d) / \tau_T\big)}
                          {\sum_{d' \in \mathcal{C}_i}\exp\!\big(f_Q^{T}(q_i)^{\top} f_D^{T}(d') / \tau_T\big)},
\]
\[
p_i^{S}(d) \;=\; \frac{\exp\!\big(f_Q^{S}(q_i)^{\top} f_D^{T}(d) / \tau_S\big)}
                          {\sum_{d' \in \mathcal{C}_i}\exp\!\big(f_Q^{S}(q_i)^{\top} f_D^{T}(d') / \tau_S\big)},
\]
and the loss minimizes the KL divergence from the (fixed) teacher distribution
to the student,
\[
\mathcal{L}_{\text{KALM}}
\;=\; \frac{1}{B}\sum_{i=1}^{B}\,
\mathrm{KL}\!\left(p_i^{T} \,\Big\|\, p_i^{S}\right)
\;=\; -\frac{1}{B}\sum_{i=1}^{B}\!\!\sum_{d \in \mathcal{C}_i}\!\!
p_i^{T}(d)\,\log p_i^{S}(d) \;+\; \Lambda,
\]
where $\Lambda$ is a constant because the teacher term $p_i^{T}$ has no gradient, and $\tau_T$ and $\tau_S$ are the teacher and student temperatures, respectively, and are set to $\tau_T = \tau_S = 0.05$ following~\cite{zhao2025kalm}. We use a batch size $B{=}1024$ so that each query sees a sufficiently rich set of hard negatives. Note that, unlike the original KaLM-Embedding-V2 recipe which jointly updates both encoders, in our setting the document encoder $f_D^{T}$ is held fixed: this is what makes the resulting student directly compatible with the Phase~1 document encoder.


\subsection{Kernel-based Unsupervised Embedding Alignment}
\label{app:kuea_loss}

Kernel-based Unsupervised Embedding Alignment (KUEA)~\cite{pmlr-v267-gong25b} asks the student to reproduce the
\emph{pairwise query-side similarity structure} induced by the teacher over a
mini-batch, without involving documents at all. For a mini-batch of queries
$\{q_i\}_{i=1}^{B}$ and a kernel $k(\cdot,\cdot)$,
\[
\mathcal{L}_{\text{KUEA}}
\;=\; \frac{1}{B(B-1)} \!\!\sum_{i \neq j}\!\!
\Big( k\!\left(f_Q^{S}(q_i), f_Q^{S}(q_j)\right)
    - k\!\left(f_Q^{T}(q_i), f_Q^{T}(q_j)\right) \Big)^{2}.
\]
The teacher kernel matrix is held fixed (no gradient), and self-similarity
terms ($i{=}j$) are excluded since they are constant. We use a polynomial
kernel of degree $d{=}3$ applied directly to $\ell_2$-normalized embeddings,
$k(u,v) = (\hat{u}^{\top}\hat{v} + 1)^{d}$, which bounds kernel values in
$[0,8]$ for both teacher and student. The original KUEA
formulation~\cite{pmlr-v267-gong25b} normalizes each kernel matrix post-hoc
(Frobenius or centered~\cite{pmlr-v97-kornblith19a}) to absorb scale and
dimension mismatch across unrelated encoders; since we feed already-normalized
embeddings of matched dimension ($d{=}128$) into the kernel, this is
unnecessary in our intra-family setting (Qwen3-Embedding-4B teacher,
Qwen3-Embedding-0.6B student).

\paragraph{Inference-time rotation alignment.}
Because KUEA only constrains \emph{pairwise} similarities among queries, its
solution is identifiable only up to an orthogonal transformation of the
embedding space: any rotation $R \in \mathbb{R}^{d \times d}$ with $R^{\top}R = I$
applied to $f_Q^{S}$ leaves $\mathcal{L}_{\text{KUEA}}$ unchanged but
arbitrarily misaligns the student query embeddings with the frozen document
embeddings $f_D^{T}$, breaking compatibility with the precomputed offline
index. To recover index compatibility without retraining or recomputing
$f_D^{T}$, we estimate a single rotation on a held-out validation set of
queries $\mathcal{V} = \{q_v\}_{v=1}^{N}$ by aligning student query
embeddings to the corresponding teacher query embeddings via the orthogonal
Procrustes problem:
\[
R^{\star} \;=\; \arg\min_{R^{\top}R = I}\;
\sum_{q_v \in \mathcal{V}} \big\| R\, f_Q^{S}(q_v) - f_Q^{T}(q_v) \big\|_{2}^{2}
\;=\; U V^{\top},
\]
where $U \Sigma V^{\top}$ is the SVD of
$\sum_{v} f_Q^{T}(q_v) f_Q^{S}(q_v)^{\top}$. At test time we apply the
\emph{same} $R^{\star}$ to every test-query embedding and retrieve against the
cached $f_D^{T}$ index, i.e. the scoring function becomes
$s(q, d) = (R^{\star} f_Q^{S}(q))^{\top} f_D^{T}(d)$. This adds a single
$d \times d$ matmul ($d{=}128$) to query-side serving cost and, crucially,
requires no changes to the offline document encoder. The numbers
reported for KUEA in Table~\ref{tab:align-ablations} use this Procrustes
post-processing.



\end{document}